# Temporally modulated phase retrieval method for weak temporal phase measurement of laser pulses

Zhi Qiao [a, b,*], Yudong Yao [a, b], Xiaochao Wang [a], Yuanyuan Jing [a, b], Wei Fan [a, *], Zunqi Lin [a]

[a] National Laboratory on High Power Laser and Physics, Shanghai Institute of Optics and Fine Mechanics, Chinese Academy of Science, Shanghai 201800, China

[b] University of the Chinese Academy of Sciences, Beijing 100049, China

Corresponding author: qiaozhi@siom.ac.cn (Zhi Qiao); fanweil@siom.ac.cn (Wei Fan)

**Abstract**: The measurement of weak temporal phase for picosecond and nanosecond laser pulses is important but quite difficult. We propose a simple iterative algorithm, which is based on a temporally movable phase modulation process, to retrieve the weak temporal phase of laser pulses. This unambiguous method can achieve a high accuracy and simultaneously measure the weak temporal phase and temporal profile of pulses, which are almost transform-limited. Detailed analysis shows that this iterative method has valuable potential applications in the characterization of pulses with weak temporal phase.

## 1. INTRODUCTION

The development of ultrashort pulses gave rise to the problem of ultrashort pulses measurements during the past two decades. Several methods such as optical autocorrelation, frequency-resolved optical gating (FROG), spectral phase interferometry for direct electric-field reconstruction (SPIDER), self-referenced spectral interferometry (SRSI), and temporal interference (TI) have been introduced to measure both the amplitude and phase of ultrashort pulses [1-5]. Nevertheless, the measurement of weak temporal phase for relatively long pulses ranging from tens of picoseconds to several nanoseconds is still difficult compared with the development of measurements for ultrashort pulses. Similar to ultrashort pulses, the temporal profile of pulses that exceed the bandwidth of available photodetectors and oscilloscopes, such as tens of picoseconds pulses, is still too fast to be resolved. The relatively low peak power is another obstacle for methods based on nonlinear effects. In addition, the weak temporal phase is even more difficult to characterize due to the narrow spectrum which is usually below 1nm.

The temporal phase of a laser pulse is quite important for high power amplifiers and lasers to optimize the structure to achieve best performance in a large laser system. Especially for large laser facilities such as laser ignition confined fusion systems (NIF, LMJ and SG series), it is extremely important to obtain the complete information, temporal profile and phase, for picosecond and nanosecond pulses during transmittance through a number of amplifiers. In the inertial confinement fusion (ICF) process, the temporal nonlinear phase accumulated in amplifiers will significantly affect the performance in the fast ignition process for picosecond pulses, and the laser fusion process for nanosecond pulses with a spectral width below 1nm [6-7]. Nevertheless, the temporal phase, which is usually below several π, is too weak to be resolved by SI, TI, FROG and linear spectrogram methods due to the narrow spectrum and low peak powers for relatively long pulses [8-11].

In order to characterize the temporal phase of pulses, two methods, the temporal transport-of-intensity equation (TIE) and temporal Gerchberg–Saxton (GS) algorithm, are proposed [12-15]. Compared with methods based on nonlinear phenomena, the temporal TIE method is a non-iterative technique that only needs two temporal intensities in the Fresnel dispersion region with equal negative and positive dispersions respectively [16]. However, the suitable dispersion condition is hard to obtain at the 1um wavelength. Besides the critical requirement for the dispersion, the temporal TIE method is also quite sensitive to noise, which limits its actual application. In contrast to the temporal TIE, the temporal GS algorithm is an iterative method, and the temporal phase can be recovered simply by using two temporal profiles before and after the dispersion. Experimental results show that the temporal GS algorithm is an effective and simple technique to retrieve the temporal phase of picosecond pulses. However the dispersion of the temporal GS method is a key factor to successfully recover the pulse phase. For pulses with a narrow spectrum, the required dispersion is impractical and noise or improper parameters will cause non-convergence and ambiguity in the GS algorithm.

Starting recently, the coherent diffraction imaging (CDI) in space has developed quickly, and various methods have been proposed to measure the spatial phase. The ptychographical iteration engine (PIE) is a successful algorithm to correctly characterize the spatial phase [17-20]. According to the duality between the space and time domain, we demonstrate a temporal iterative algorithm, similar to the ptychographical algorithm, based on the temporally movable phase modulation process using a commercial electro-optic phase modulator. Compared with other methods, this temporally modulated phase retrieval (TMPR) algorithm can measure the weak phase of pulses with a narrow spectrum. Due to its temporally movable phase modulation process, the TMPR algorithm can achieve higher resolution and sensitivity than other methods such as GS algorithm. The TMPR method is also robust to noise and measurement error, which makes it suitable in laser systems with a series of amplifiers with low optical signal-to-noise ratio (OSNR). Another advantage of TMPR is the ability to measure the temporal phase and temporal profile simultaneously, which is also convenient to completely characterize the laser pulses.

## 2. TEMPORALLY MODULATED PHASE RETRIEVAL METHOD

### 2.1 Theory of TMPR algorithm

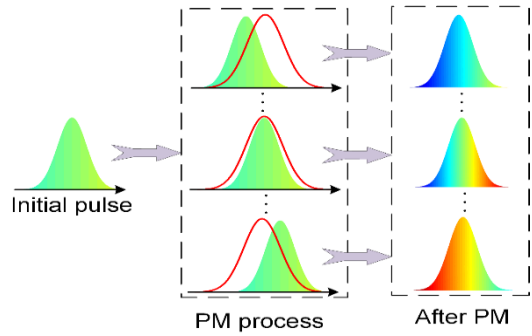

Fig. 1 Illustration of the temporal phase modulation process in the TMPR algorithm (red line: modulating function; PM process: phase modulation process).

In the TMPR algorithm, a modulating function, which can be temporally shifted relatively to the laser pulse, is necessary to modulate the initial pulses. Theoretically, the modulating effect can be either the phase modulation or the amplitude modulation. Because we intend to measure the weak temporal phase, the phase modulation process is applied and an illustration of the TMPR process is shown in Fig. 1. The initial pulse is $u(t)$ and the modulating function is $m(t)$. As shown in Fig. 1, the initial signal pulse is phase modulated temporally by the modulating function and can be written as

$$A_{t_0+t_j}^{ini}(t) = u_{t_0}(t) m_{t_0+t_j}(t),\qquad(1)$$

where $m_{t_0+t_j}(t)$ is the modulating function shifted in time relatively to the initial pulse with the $j$-th position $t_j$, and $t_0$ is the central position. The complex amplitude $A_{t_0+t_j}^{ini}$ after phase modulation cannot be acquired directly, but its temporal profile or spectrum is available. The measurements of the temporal profile after transmitting in a dispersion fiber can be utilized to refresh the complex amplitude, which is

$$A_{t_0+t_j}^{r}(t) = \frac{A_{t_0+t_j}^{D}(t)}{\left|A_{t_0+t_j}^{D}(t)\right|}\sqrt{I_{t_0+t_j}(t)}, \tag{2}$$

where $I_{t_0+t_j}(t)$ is the measured temporal intensity and $A_{t_0+t_j}^{D}$ is the laser pulse after the dispersion process.

In addition to the temporal profile, the spectrum is also sufficient to obtain a modified electric field

$$A_{t_0+t_j}^{r}(t) = FT^{-1}\left\{\frac{FT\{A_{t_0+t_j}^{ini}(t)\}}{\left|FT\{A_{t_0+t_j}^{ini}(t)\}\right|}\sqrt{S_{t_0+t_j}(\omega)}\right\}, \tag{3}$$

where $S_{t_0+t_j}(\omega)$ is the measured spectrum and $FT$ denotes the Fourier transformation. According to the relationship between the initial complex amplitude and the refreshed complex amplitude given in Eq. (2) and Eq. (3), we can define a least-square estimation of the unknown initial laser through the minimization of the criterion

$$H = \sum_j \left\{ \sum_t \left| u'_{t_0} m_{t_0+t_j} - A_{t_0+t_j}^{r} \right|^2 + \sum_t \gamma_{t_0} \left| u'_{t_0} - u_{t_0} \right|^2 \right\} = \sum_j H_j, \tag{4}$$

where $u'_{t_0}$ is the modified signal laser and $\gamma_{t_0}$ is the weight of the effect of the modified signal laser on the criterion.

Similar to the well-known PIE algorithm [18], we can calculate the gradient of $H_j$ with respect to $u'_{t_0}$

$$\begin{aligned}\frac{\partial H_j}{\partial u'_{t_0}} &= \sum_t \{[m^*_{t_0+t_j}(u'_{t_0} m_{t_0+t_j} - A_{t_0+t_j}^{r}) + \gamma_{t_0}(u'_{t_0} - u_{t_0})] \\ &\quad + [m^*_{t_0+t_j}(u'_{t_0} m_{t_0+t_j} - A_{t_0+t_j}^{r}) + \gamma_{t_0}(u'_{t_0} - u_{t_0})]^*\}.\end{aligned} \tag{5}$$

When the least-square estimation of the unknown signal pulse is at the minimal point, it is reasonable to consider that the modified complex amplitude $u'_{t_0}$ converges to $u(t)$. Therefore if the gradient of $H_j$ is zero, we can deduce the updating function

$$u'_{t_0} = u_{t_0} + \frac{m^*_{t_0+t_j}}{\left|m_{t_0+t_j}\right|^2 + \gamma_{t_0}}\left(A_{t_0+t_j}^{r} - m_{t_0+t_j} u_{t_0}\right). \tag{6}$$

The value $\gamma_{t_0}$ can be a constant coefficient to control the effect of the modified laser pulse on $H$. However $\gamma_{t_0}$ must be nonzero to prevent the divide-by-zero occurring if $|m(t)| = 0$. Considering the realization of this algorithm, a parameter $\kappa$ ($\kappa > 0$) is introduced to control the speed of convergence and the updating function is redefined as

$$u'_{t_0} = u_{t_0} + \kappa \frac{m^*_{t_0+t_j}}{\left|m_{t_0+t_j}\right|^2 + \gamma_{t_0}}\left(A_{t_0+t_j}^{r} - m_{t_0+t_j} u_{t_0}\right). \tag{7}$$

**2.2 Procedure of TMPR algorithm**

Here we describe the detailed procedure to retrieve the temporal phase through the TMPR algorithm. Before starting the TMPR procedure, a series of measurements, which can be post-dispersion temporal profiles or spectra, must be prepared. We take spectral measurements as an example, because the measurement of a spectrum is simpler than that of a temporal profile for which a certain amount of dispersion is necessary to transform the phase information into the time domain. Additionally, the process to measure spectrum would not be restricted by the analog bandwidth of electrical elements with respect to oscilloscopes.

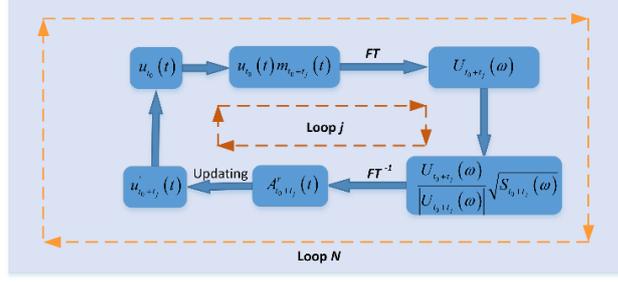

Fig. 2 Schematic of the TMPR algorithm for temporal phase retrieval.

The TMPR temporal phase retrieval process is shown in Fig. 2 where the updating process denotes the signal laser refreshing method as shown in Eq. (7). The TMPR contains two circulations, which are Loop $j$ and Loop $N$. Loop $j$ is a circulation to refresh the unknown signal laser $u_{t_0}$ by the whole measured spectra with different time shifting $t_j$ of the modulation function and loop $N$ is the total iteration number to obtain a converged solution. Because the phase modulation is chosen as the modulation function, $m(t)$ can be written as

$$m_{t_0+t_j}(t) = e^{i\left(\pi V(t_0+t_j)/V_\pi + \varphi_0\right)} \tag{8}$$

where $V_\pi$ is the half-wave voltage of the phase modulator and $V(t_0 + t_j)$ represents the electric signal acting on the phase modulator. Due to the inner imperfect quality of the electro-optic phase modulator, a constant phase $\varphi_0$ will also affect the phase, but this small constant phase can be neglected, compared with the active phase modulation signal.

The detailed steps of the TMPR algorithm are demonstrated as follows:

(1) Produce an initial guess $u_{t_0}$ as the initial pulse to start the phase retrieval procedure. The initial guess can be a totally random noise function, which is not a critical parameter to successfully obtain a stable solution.

(2) The signal pulse is then phase modulated by the electric signal with a time shifting of $t_j$ and can be calculated with Eq. (1).

(3) The spectrum $S_{t_0+t_j}(\omega)$ measured by a spectrometer is utilized to refresh the complex amplitude after phase modulation as shown in Eq. (3) and the refreshed $A^r_{t_0+t_j}$ is obtained. The measured spectra are actually the only boundary condition in the TMPR algorithm.

(4) When the refreshing process is complete, the guess solution $u_{t_0}$ can be updating by the refreshed complex amplitude $A^r_{t_0+t_j}$ using Eq. (7). Parameter $\kappa$ can control the feedback which affects the convergence speed in this algorithm.

(5) Repeat steps from (2) to (4) for the modulating signal with the next ($j$+1) th temporal shift until all the measured spectrum has been used.

(6) Repeat the steps from (2) to (5) until the estimated error achieves a stable state or the iteration number exceeds the maximum limit. The error of this retrieval algorithm is defined as

$$\varepsilon_{error} = \frac{\sum_j \int_t \left(\left|FT\left\{A^r_{t_0+t_j}\right\}\right| - \sqrt{S_{t_0+t_j}(\omega)}\right)d\omega}{\sum_j \int_t \sqrt{S_{t_0+t_j}(\omega)}d\omega}. \tag{9}$$

Because the spectrum $S_{t_0+t_j}(\omega)$ after phase modulation should be able to be measured by a commercial spectrometer, the modulating function must be strong enough to broaden the spectra sufficiently to retrieve the signal laser successfully. In another word, the spectral width after phase modulation should be much wider than the resolution of a spectrometer, which can be written approximately as:

$$\Delta\Omega < \frac{1}{\beta}\varphi' \approx \frac{\Delta\varphi}{\beta\Delta t}, \quad (10)$$

where $\Delta\Omega$ is the angular frequency resolution of spectrometer, $\Delta\varphi$ is the variation of phase during time range of $\Delta t$ and $\beta$ is a coefficient which determines whether the variation of phase is enough. For simplicity, a normal Gaussian function is chosen as the modulating signal so that the $\Delta\varphi$ needed for a pulse to be measured can be written as

$$\varphi_M > \frac{2\pi c \beta \Delta T \Delta\lambda}{\lambda^2}, \quad (11)$$

where the $\varphi_M$ is the peak phase of modulating signal, $\Delta T$ is the pulse width of modulating signal and $\Delta\lambda$ is the resolution of spectrometer.

Generally the coefficient $\beta$ can be chosen as 3 and a perfect performance can be obtained in our experience. It should also be noted that Eq. (11) is only suitable for the situation where the duration of modulating signal is larger than the pulse to be measured.

## 3. SIMULATION AND ANALYSIS

We consider the typical situation where the phase of the laser pulses is induced by self-phase modulation (SPM), which is the main cause of nonlinear phase in most cases. A short pulse with a duration of 30ps is used as the signal laser, of which the nonlinear phase induced by SPM is $\alpha\pi$. Although a picosecond pulse is taken as the signal laser, it should be noted that this algorithm is also suitable to measure the temporal phase of nanosecond pulses. Considering the limited bandwidth of analog electric elements, the Gaussian modulating function with a duration of 150ps is used to modulate the signal laser. If the resolution of a commercial spectrometer is 0.01nm, the modulation depth of the phase modulating function is chosen as 4 which is not a problem for commercial electric elements and phase modulators. Besides the modulation depth and duration of the modulating function, the time shifting is also an important process to retrieve the temporal phase. For the simplicity, we use the equal interval shifting method with a time shifting of $(1-\eta)\tau$, where $\tau$ is the duration of the signal pulse and $\eta$ is the overlapping ratio during the time shifting. To fully make use of the modulating signal, the signal laser should be shifted at least in a range of the duration of the modulating function. The noise of the signal laser and noise in the measurement of spectra are also taken into consideration and normally the random noise levels are both 1%.

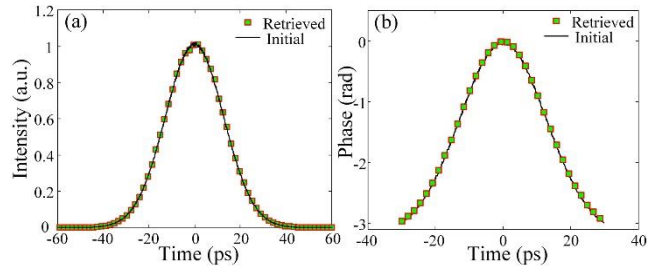

Fig. 3 Retrieval results after 200 iterations. (a)Temporal profile of the initial pulse and retrieval pulse; (b) temporal phase of the initial pulse and retrieval pulse.

Through the TMPR procedure shown above, the retrieval results, when $\alpha$ equals 1 and $\eta$ is 0.6, are shown in Fig. 3 after 200 iterations. It can be seen that the temporal phase and temporal profile can both be recovered successfully with negligible retrieval error. During the iteration, it is interesting that the temporal phase converges faster and is more accurate than the temporal profile. Even when the temporal profile cannot be recovered correctly, the retrieval result of the temporal phase still shows great accuracy, which is consistent with spatial phase retrieval algorithms such as GS and PIE. Considering that the spectrum is more sensitive to the temporal phase compared with the temporal profile, it is reasonable that the retrieval of the temporal phase is simpler. Compared with other methods, the TMPR algorithm can simultaneously recover the temporal phase and the temporal profile, which is convenient and accurate due to the mutual restrictive relationship between the temporal phase and temporal profile with certain measured spectra in the iterative algorithm.

The overlapping area is the key to successfully retrieve the spatial phase in the spatial phase retrieval algorithm, PIE. Therefore it is intuitive to think that the overlapping area is also vital in this iterative algorithm. However, comparing the retrieval results with different overlapping ratios shown in Fig. 4, it proves that the overlapping ratio is not necessary. A large overlapping ratio can accelerate the convergence but after enough iterations, the final retrieval result shows no difference as shown in Fig. 4. Even if the overlapping area is zero which indicates that the signal lasers after time shifting are completely non-overlapping with each other, the recovered temporal phase and profile are still same with the overlapping result. The feature of the TMPR algorithm that the overlapping area is unnecessary indicates that the TMPR method is different from the spatial PIE phase retrieval algorithm. This temporal phase retrieval method is more like the spatial phase retrieval method based on the structural illumination.

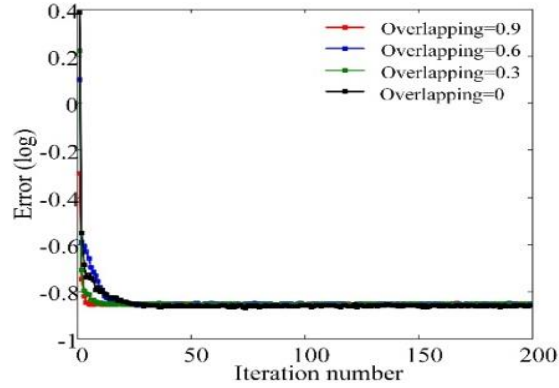

Fig. 4 Retrieval error after 200 iterations with different overlapping ratios.

The TMPR algorithm is based on the active phase modulation process which determines that the TMPR method is quite suitable to measure the weak nonlinear temporal phase. The limitation of the minimal temporal phase that can be measured is studied with same parameters as in Fig. 3 except for the initial nonlinear phase of the signal laser. Retrieved temporal phases are shown in Fig. 5 with initial nonlinear phase of the signal laser of $0.05\pi$, $0.01\pi$, $0.005\pi$ and $0.001\pi$ respectively. It can be seen that the performance of the phase retrieval abates with weaker signal phase but the temporal phase of $0.005\pi$ can still be recovered successfully as shown in Fig. 5 (c). If the signal phase is lower than $0.001\pi$, the retrieved phase experiences large error. Nevertheless even $0.01\pi$ is sufficiently weak in laser systems that need to take the nonlinear phase of laser pulses into account. The TMPR algorithm shows great advantage in the measurement of the ultra-weak temporal phase. Additionally, the minimal temporal phase that can be measured is related with the modulation depth and if a phase modulation process with large modulation depth is applied, the lower temporal phase of the signal laser can be retrieved successfully.

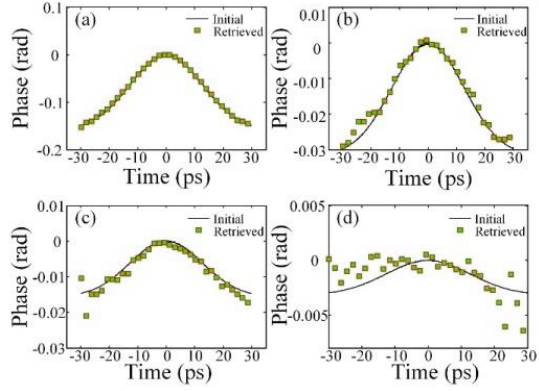

Fig. 5 Influence of the weak signal phase on the retrieval phase of TMPR. (a) 0.05π; (b) 0.01π; (c) 0.005π; (d) 0.001π.

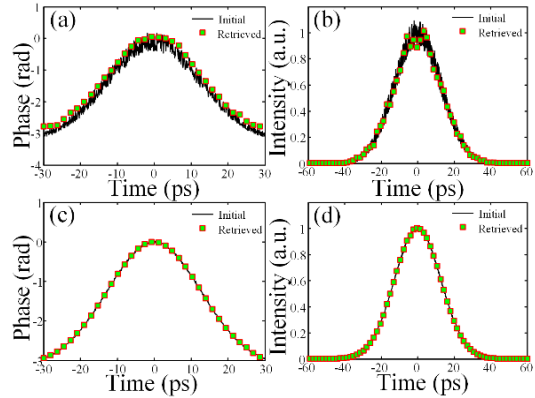

Fig. 6 Retrieval results after 200 iterations with noise in the signal laser. (a), (b) Phase and temporal profile with a signal noise of 10%; (c), (d) phase and temporal profile with signal noise of 1%

The noise has a great impact on the temporal GS algorithm and temporal TIE method that a slight perturbation will result in a completely wrong solution. To compare the effect of noise on the TMPR algorithm, we ran a series of conditions with different noise levels of the signal pulse as well as the measured spectrum. Figure 6 is the retrieved phase and intensity under random signal noises of 10% and 1% with the same parameters in Fig. 3. As Fig. 6 (a) and (b) show, the phase and profile of the initial signal laser contain a large noisy fluctuation. However, the retrieved phase and temporal profile still show a great accuracy. With a lower noise level of 1%, the retrieved results are nearly perfect, which indicates that the TMPR algorithm is robust to the signal noise.

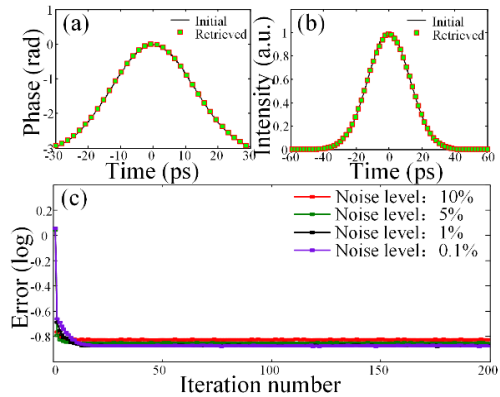

Fig. 7 Retrieval results after 200 iterations with noise in spectral measurements. (a), (b) Phase and temporal profile with a noise level of 10%; (c) retrieval error under different noise levels of spectral measurements.

For the measurement of the temporal phase in a large laser system, the systematic error is influenced by the signal noise as well as the noise of the spectral measurement, which is restricted by instruments of spectrometer and the instability of the laser system. Just like the analysis of the signal noise, the influence of the noise in spectral measurements is also studied and the retrieval phase is shown in Fig. 7. The noise in spectral measurements is simulated by adding a random fluctuation on the measured spectra. As the retrieval error curves with spectral noise of 10%, 5%, 1% and 0.1% respectively shown in Fig. 7 (c), the noise in spectral measurement only slightly affects the retrieval error after 200 iterations. Moreover even under a noise level of 10%, the temporal phase and profile shown in Fig. 7 (a) and (b) are correctly recovered. Theoretically, due to the random nature, the noise in a certain spectrum could be remedied by other spectra with different time shifting. As a result of the iterative character and time shifting process, the TMPR algorithm can converge to a solution with minimal error to the measured spectra, which determines that this algorithm can tolerate large noise in spectral measurements. Combining this with the numerical analysis above, it can be concluded that the TMPR algorithm is robust against the systematic noise, which is an excellent advantage for the measurement of the temporal phase in a complex and restricted laser system.

## 4. CONCLUSIONS

In conclusion, we propose a temporally modulated phase retrieval (TMPR) algorithm that is based on the duality between the time and space domain. The TMPR algorithm is an iterative algorithm similar to the successful spatial phase retrieval PIE method. However the simulation results show that the TMPR algorithm has little dependence on the overlapping area which is vital for the spatial PIE based algorithm. Besides the temporal phase, the temporal profile can also be retrieved simultaneously, which is convenient in actual applications. Due to the temporal movable phase modulation process, the TMPR algorithm performs excellent for the measurement of weak temporal phase. Even for pulses with a nonlinear phase of $0.005\pi$, the temporal phase and temporal profile can both be recovered successfully. The effect of noise on the retrieval results is analyzed and simulations show that the TMPR algorithm is robust against the systematic noise which is quite important for complex laser systems. The TMPR algorithm together with the measurement schema provides new perspectives to measure the weak temporal phase in complex high power laser systems.

**ACKNOWLEDGEMENTS**

The work reported in this paper is supported by the National Natural Science Foundation of China (61205103).